\documentclass[10pt]{iopart}
\usepackage{iopams,graphicx,subfigure,braket,mathptmx,color}

\newcommand{\openone}{\leavevmode\hbox{\small1\normalsize\kern-.33em1}}
\newcommand{\Var}{\mathop{\mathrm{Var}} \nolimits}

\begin{document}

\title{Axial superlocalization with vortex beams}

\author{D~Koutn\'{y}$^{1}$, Z~Hradil$^{1}$, 
J~\v{R}eh\'{a}\v{c}ek$^{1}$ and L~L~S\'{a}nchez-Soto$^{2,3}$}

\address{Department of Optics, Palack\'{y} University, 17. listopadu 12, 77146 Olomouc, Czech Republic}

\address{Departamento de \'{O}ptica, Facultad de F\'{i}sica, Universidad Computense, 28040 Madrid, Spain}

\address{Max-Planck-Institut f\"{u}r die Physik des Lichts, Staudtstra{\ss}e 2, 91258 Erlangen, Germany}

\ead{rehacek@optics.upol.cz}
\vspace{10pt}
\begin{indented}
\item[\today]
\end{indented}


\begin{abstract}
Improving axial resolution is of paramount importance for three-dimensional optical imaging systems. Here, we investigate the ultimate precision in axial localization using vortex beams. For Laguerre-Gauss beams, this limit can be achieved with just an intensity scan. The same is not true for superpositions of Laguerre-Gauss beams, in particular for those with intensity profiles that rotate on defocusing. Microscopy methods based on rotating vortex beams may thus benefit from replacing traditional intensity sensors with advanced mode-sorting techniques.
\end{abstract}

\vspace{2pc}


\eqnobysec


\section{Introduction}

Discerning the nanometer‐scale details of living cells, tissues, and materials is of utmost importance for many modern research efforts.  A trail towards to this holy grail was opened with the appearance of a set of methods, dubbed with the generic term of superresolution microscopy~\cite{Natsupres:2009aa,Schermelleh:2019aa}, capable of bypassing the diffraction limit~\cite{Abbe:1873aa,Rayleigh:1879aa,Ram:2006aa}: a barrier that was traditionally thought to be impenetrable. 

A number of these techniques can also reveal three-dimensional (3D) structural details: relevant examples include stimulated-emission-depletion microscopy~\cite{Hell:1994ab}, PSF engineering~\cite{Huang:2008aa,Pavani:2009aa,Jia:2014aa,Tamburini:2006aa,Paur:2018aa}, photoactivated-localization microscopy~\cite{Betzig:2006aa}, and multiplane detection~\cite{Juette:2008aa,Dalgarno:2010aa,Abrahamsson:2012aa}, to cite only but a few.  All of them rely on a very accurate localization of point sources; they differ in how point objects are excited and how the corresponding emitted photons are collected.

For 3D imaging, the emitter is fluorescently labeled and determining its axial position is an indispensable ingredient. This problem has been  throughly examined and some impressive results have been demonstrated so far~\cite{Diezmann:2017aa}.  However, the fundamental depth precision achievable by any such engineering method has been considered only lately~\cite{Tsang:2015aa,Nair:2016ab,Backlund:2018aa}. The rationale behind is to make a systematic use of the quantum Fisher information (QFI)~\cite{Petz:2011aa} and the associated quantum Cram\'er-Rao bound (QCRB) to get a measurement-independent limit~\cite{Helstrom:1976ij,Holevo:2003fv}. This is much along the lines of the work of Tsang and coworkers to quantify the transverse two-point resolution~\cite{Tsang:2016aa,Lupo:2016aa,Nair:2016aa,Ang:2016aa,Tsang:2017aa}, which has led to the dispelling of the Rayleigh curse~\cite{Paur:2016aa,Yang:2016aa,Tham:2016aa,Yang:2017aa}.

In a recent work~\cite{Rehacek:2019aa}, the ultimate precision {in axial localization} using Gaussian beams has been established.  This limit can be attained with just one intensity scan, as long as the detection plane is placed at one optimal position. 

In this paper, we generalize those results and derive quantum limits for axial localization with Laguerre-Gauss (LG) beams, which carry quantized orbital angular momentum~\cite{Allen:1992aa}. 
{Here, the beam waist acts as a realization of the light emitted by a point source after e.g. mode conversion. Another relevant situation is the reflection of the beam from a surface in surface topology measurements, etc.} By linearly superposing  different LG modes, one can realize beams with amplitude, phase, and intensity patterns that simply rotate, under free space propagation, maintaining the transverse shape. These rotating structures lie at the core of a variety of sensing techniques~\cite{Piestun:2000aa,Greengard:2006aa,Prasad:2013aa,Baranek:2015aa}. 

We  demonstrate that a meager part of the full (quantum) information is available in intensity scans and only a small fraction of this can be attributed to the rotation. This clearly confirms the potential of modal expansions inspired by quantum information protocols~\cite{Fabre:2020aa}, which allow for reaching the QCRB and thus can be considered as the optimal measurement. Our results make 3D superresolution imaging  more feasible and potentially useful for improving the resolution of optical microscopes.

\section{Theoretical model}

The problem we address here is to estimate the distance traveled by a vortex beam from the beam waist to an arbitrary detection plane. We thus consider the beam waist as an object whose axial distance is to be determined. In what follows, we shall represent the fields using Dirac notation, for it makes it straightforward to expand the theory to other types of light states.  

We take the beam to be represented by the pure state $|\Psi (0) \rangle$, where $z=0$ denotes the position of the object plane. The axial displacement is thus characterized  by a unitary operation 
\begin{equation}
\label{eq:unitpar}
|\Psi (z)\rangle =  \e^{\rmi G \,  z} \; | \Psi (0) \rangle \, ,
\end{equation} 
where the Hermitian operator $G$ is the generator. To pinpoint the specific form of $G$, it is appropriate to use the transverse-position representation  $\Psi (x,y; z) = \langle x,y | \Psi (z) \rangle$. It then follows directly from equation~(\ref{eq:unitpar}) that
\begin{equation}
 \rmi  \partial_{z} \Psi (x,y; z) = -  G \,  \Psi (x,y; z) \, .
\end{equation}
On the other hand, vortex beams are solutions of the paraxial Helmholtz equation~\cite{Franke-Arnold:2008sw,Barnett:2017aa} 
\begin{equation}
\label{eq:Parax}
2\rmi k \partial_z \Psi(x,y,z) = \nabla_{T}^{2}\Psi(x,y,z) \, ,
\end{equation}
where  $k$ is the wavenumber and $\nabla^{2}_{T} = \partial_{xx} + \partial_{yy}$ is the transverse Laplacian. A direct comparison leads us to 
\begin{equation}
\label{eq:Gdelta}
G \mapsto - \frac{1}{2k} \nabla^{2}_{T} \, .
\end{equation}

The detection plane is placed at $z$, wherein we perform an arbitrary measurement.  {Given the formal analogies between spatial modes in wave optics and pure states in quantum theory, and also the mathematical similarities in describing evolution and detection of such objects, the amount of information about the axial distance $z$ carried by the measured signal is quantified by the QFI. For pure states, as it is our case, the QFI reduces to~\cite{Helstrom:1976ij}} 
\begin{equation}
 \mathcal{Q}( z ) = 4 \Var (G) \, . 
\end{equation} 
Except for the factor 4, the QFI is the variance of the generator $G$ computed in the initial state $| \Psi(0) \rangle$. According to the time-honored QCRB, the variance of any unbiased estimator $\widehat{z}$ of the axial distance $z$ satisfies
\begin{equation}
\Var (\widehat{z}) \ge \frac{1}{\mathcal{Q} (z)} \, ,
\end{equation}
whose saturation provides the ultimate precision in axial distance estimation.

To proceed further, we take the structure of the transverse field to correspond with LG modes 
\begin{eqnarray}
\label{eq:LGDef}
& \displaystyle
\mathrm{LG}_{pl}(r,\phi,z)  = \langle r,\phi,z|p,l \rangle = 
\sqrt{\frac{2p!}{\pi (p+|l|)!}} \frac{1}{w(z)}\left [\frac{\sqrt{2}r}{w(z)}\right]^{|l|} & \nonumber \\
& \times \displaystyle 
L_p^{|l|}\left(\frac{2r^2}{w(z)^2}\right)  \, e^{- \frac{r^2}{w(z)^2}} \;
 \exp \left( \rmi \left [ \frac{kr^{2}}{2 R(z)} - l \phi - \psi_{pl}(z) \right ] \right) \, ,  &   
\end{eqnarray}
where $(r,\phi, z)$ are cylindrical coordinates,  $L_p^{|l|}(\cdot)$ is the generalized Laguerre polynomial, $l \in \{0,\pm 1,\pm 2, \dots\}$ is the azimuthal mode index and $p \in \{0, 1, 2,\dots\}$ is the radial index. The parameters $R(z)$, $w(z)$, and $\psi_{pl}(z)$ are 
\begin{eqnarray}
\label{eq:ParDef}
R(z) & = & z\left[1+(z_R/ z)^2\right], \nonumber \\
w^2(z) & = & w^2_0\left[1+\left(z/z_R \right)^2 \right], \\
\psi_{pl} (z) &= &(2p+|l|+1)\arctan(z/z_R), \nonumber
\end{eqnarray}
and represent the radius of curvature of the wave front, the beam radius, and the Gouy phase~\cite{Feng:2001aa}, respectively, at an axial distance $z$ from the beam waist located at $z=0$. Here, $z_{\mathrm{R}} = k w_0^2/2$ is the Rayleigh length and $w_{0}$ the beam waist radius~\cite{Siegman:1986aa}.

\section{Quantum limit for axial localization with vortex beams}

To facilitate the derivation of the QFI corresponding to axial displacements of vortex beams it is advantageous to use an established correspondence between eigenstates of a two-dimensional harmonic oscillator and paraxial beams~\cite{Nienhuis:2017aa}. For the case of LG beams, one defines 
\begin{equation}
a_{\pm} = \frac{1}{\sqrt{2}} ( a_{\xi} \mp i a_{\eta} ) \, ,
\end{equation}
where $a_{\xi}$ and $a_{\eta}$ are dimensionless bosonic operators for each independent amplitude of the oscillator and obey the standard commutation relations $[a_{s}, a^{\dagger}_{s^{\prime}} ] = \delta_{s s^{\prime}}$ ($s,s^{\prime} \in \{\xi , \eta \}$). Similar relations are obeyed by $a_{\pm}$.

The eigenstates of the harmonic oscillator $|n_{+}, n_{-} \rangle$, generated by the action of $a_{+}^{\dagger}$ and  $a_{-}^{\dagger}$ on the vacuum, are nothing but LG modes with azimuthal and radial indices given by 
\begin{equation}
l = n_{+} - n_{-}\, , 
\qquad 
p = \min (n_{+},n_{-}) \, .
\end{equation} 
{Applying the usual definition of the momentum operator 
$p_s =\case{1}{\sqrt{2} \rmi} (a_{s} - a^{\dagger}_{s} )$,}
we have 
\begin{equation}
\label{eq:Gen}
\widetilde\nabla^2_T = p_{\xi}^2 + p_{\eta}^2 = (p_{\xi} + i p_{\eta}) (p_{\xi} - i p_{\eta}) =  (a_{+} - a^{\dag}_{-})
(a^{\dag}_{+} -a_{-} ) \, . 
\end{equation}
If we recall (\ref{eq:Gdelta}), and take into account that, in the proper units, $(x,y) \mapsto ( \sqrt{2}\xi/w_0,\sqrt{2}\eta/w_0\})$, we get $\nabla^2_T \mapsto 2\widetilde\nabla^2_T/w_0^2$. In this way, the QFI of a pure LG mode $\ket{n_+,n_-}$ reads:   
\begin{equation}
\label{res1}
\mathcal{Q} (z) = \frac{1}{z_{\mathrm{R}}^2} (2 n_{+} n_{-}  + n_{+} + n_{-}+ 1 ) =  \frac{1}{z_{\mathrm{R}}^2} [ 2p(p+|l|)+2p+|l|+1 ] \,.
\end{equation}
For the particular case of the Gaussian mode $\mathrm{LG}_{00}$ we get
\begin{equation}
\label{eq:quant_lim}
\mathcal{Q} (z) = \frac{1}{z_{\mathrm{R}}^{2} } \, ,
\end{equation}
that is, the quantum bound (per single detection) is precisely the Rayleigh range~\cite{Zhou:2019aa}. Note that the QFI is \textit{linear} in $|l|$, which means that axial localization can be improved by using LG beams with large OAM. 

Other sets of Gaussian transverse modes can be characterized using the so-called  Hermite-Laguerre sphere~\cite{Simon:2000aa,Visser:2004aa}. These modes are represented by a point, of spherical coordinates $(\theta, \phi)$, on that sphere, and they are generated by the rotated operators
\begin{eqnarray}
\label{eq:GenCreat}
a_{1} (\theta,\phi) & = &
a_{+} e^{-i\phi/2} \cos \left (\frac{\theta}{2} \right ) + 
a_{-} e^{i\phi/2} \sin \left ( \frac{\theta}{2} \right ), 
\nonumber \\
& & \\
a_{2} (\theta,\phi) & =  &
- a_{+} e^{-i\phi/2} \sin \left ( \frac{\theta}{2} \right ) +
a_{-} e^{i\phi/2} \cos \left ( \frac{\theta}{2} \right ) \, . 
\nonumber 
\end{eqnarray}
In particular,  $\theta=\pi/2$ gives rise to Hermite-Gauss (HG) modes, and $\theta = 0, \pi$ to LG modes analyzed above. Combining equations~(\ref{eq:Gen}) and (\ref{eq:GenCreat}), we get a direct generalization of~(\ref{res1}); viz, 
\begin{equation}
\label{res2}
\fl 
\mathcal{Q} = \frac{1}{z_{\mathrm{R}}^{2}} [ 4+n_{1}+n_{2} (3+n_{2})+ 
n_{1}(3+4n_{2})+
(n_{1}-n_{1}^2+n_{2}+4n_{1}n_{2}-n_{2}^2)\cos (2\theta )]\,,
\end{equation}
where $n_{1}$ and $n_{2}$ denote the eigenvalues of the corresponding number operators. Notice that the QFI (\ref{res2}) is independent of $\phi$ and is optimized by  (\ref{res1}). This proves that LG modes are better than their HG counterparts, as axial localization is concerned.

\begin{figure}[t]
\centering
\includegraphics[height=5cm]{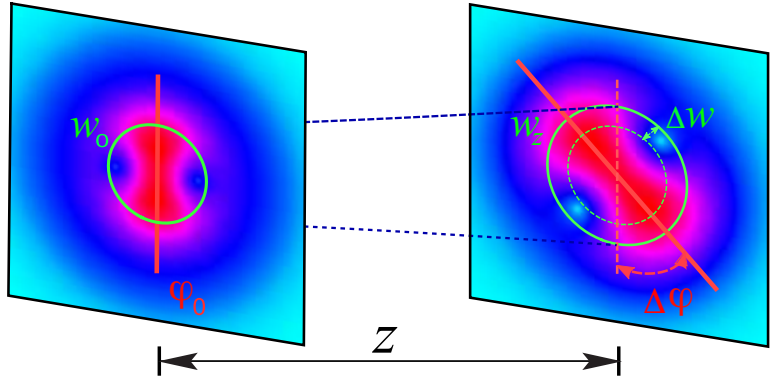}
\caption{Sketch of the evolution of the superposition of two LG modes, with $l=2$ and $l^{\prime}=0$. The interference pattern is subject to rotation as well as divergence during propagation.}
\label{fig:lg}
\end{figure}

Apart from a trivial divergence, transversal intensity profiles of pure vortex beams do not change on propagation. More complex intensity transformations can be realized by superposing two or more vortex beams. In particular, rotating structures are of interest in microscopy~\cite{Piestun:2000aa,Greengard:2006aa,Prasad:2013aa,Baranek:2015aa}. Rotation of highly symmetric spots is easily detected and the corresponding defocusing parameter can be estimated from the measured rotation angle. This is roughly sketched in figure~\ref{fig:lg}.

To simplify the details as much as possible, we take the simple example of the superposition of  two LG modes with different azimuthal numbers $l\neq l^{\prime}$, and $p=p^{\prime}=0$
\begin{equation}
|\Psi_{ll^{\prime}}\rangle = \frac{1}{\sqrt{2}}
(|\mathrm{LG}_{0l} \rangle +|\mathrm{LG}_{0l^{\prime}}\rangle ) 
= \frac{1}{\sqrt{2}} (\ket{n_+,0}+\ket{n^{\prime}_+,0} ) \, .
\end{equation}
In this case, the QFI about the axial position of the source reads
\begin{equation}
\label{res3}
\mathcal{Q} = \frac{1}{z_{\mathrm{R}}^{2}} [ 4 + 2(|l| + |l^{\prime}|) + (|l| - |l^{\prime}|)^2 ] \, .
\end{equation}
Considering OAM as a resource for axial localization and using the maximum available OAM for one of the components of the superposition, $l=\pm |l_\mathrm{max}|$, the QFI is maximized when the second component is in the fundamental mode $l^{\prime}=0$, whereupon $\mathcal{Q}_{\max} = [ 4+2|l_{\max}|+|l_{\max}|^2]/z_{\mathrm{R}}^{2}$,and the quantum bound becomes \textit{quadratic} in $|l|$. At first sight, the better performance of vortex superpositions (\ref{res3}) over pure vortex beams (\ref{res1}) is somehow related to beam rotation. This might suggest a proxy for the measured axial distance traveled from the waist to the detection plane. However, as we show in the next section, things are not that simple.

\section{Intensity detection}

In single-parameter estimation, the QFI can always be accessed and the corresponding QCRB saturated with a von Neumann measurement projecting the measured signal on the eigenstates of the symmetric logarithmic derivative of the density matrix~\cite{Helstrom:1976ij}. In our context, the practical implementation of such measurements requires a spatial mode demultiplexer/sorter that performs simultaneous projection of the signal onto a complete orthonormal set of modes~\cite{Rehacek:2017aa,Rehacek:2017ab}. Mode separation is usually achieved with a sequence of spatial light modulators implementing a suitable unitary transformation. Unavoidable systematic errors and losses introduced by such complicated experimental setups may ruin any theoretical advantage offered by optimal  strategies.

In consequence, we consider the performance of the possibly inferior, but much more robust intensity detection, because it is the simplest method at hand for the experimentalist. As the information (\ref{res1}) and (\ref{res2}) about the axial distance is carried by both intensity and phase of the measured beam, we might aptly ask how much information is sacrificed by completely ignoring the phase. 

\begin{figure}[tbp]
\centering\includegraphics[height=5.5cm]{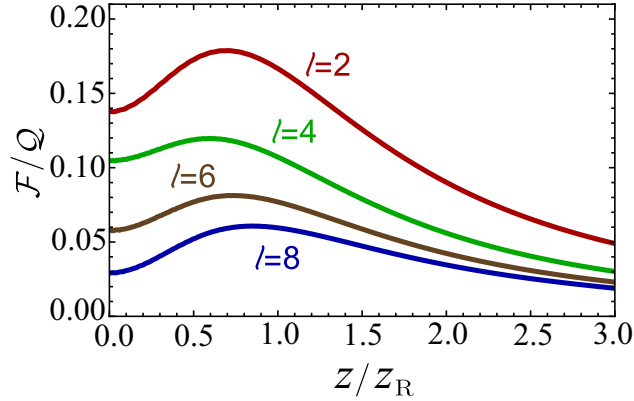}
\caption{Fisher information about axial distance (relative to the QFI)  for different superpositions $(| \mathrm{LG}_{0l} \rangle + | \mathrm{LG}_{00} \rangle)/\sqrt{2}$ as a function of the position  $z$ (relative to the beam waist) of the detection plane.} 
\label{fig:FIQl}
\end{figure}

As usual, due to the noise, the detection can be considered as a random process. In consequence, the (normalized) beam intensity, 
\begin{equation}
p(r, \phi|z)=|\Psi(r, \phi;z)|^2 \, ,
\end{equation} 
can be seen as the probability density of a detection conditional on the axial distance $z$. We take the detection as dominated by shot noise, which obeys a Poisson distribution~\cite{Fuhrmann:2004aa}: although this  neglects nonclassical effects,  it is still a suitable model for realistic microscopy.

The classical Fisher information about $z$, per single detection, thus reads~\cite{Fisher:1925aa}  
\begin{equation}
\label{eq:cfi}
\mathcal{F} = \int_0^{\infty}\int_0^{2\pi} 
\frac{\left[\partial_z p(r,\phi|z)\right]^2}{p(r,\phi|z)} \, 
r drd\phi \, ,
\end{equation}
and it is a suitable tool to quantify the information content about axial displacements accessible from the detected transversal intensity profile. For simplicity, we take the pixel size negligibly small, so that any sampling effect can be ignored. We also define the radial and azimuthal Fisher informations 
\begin{eqnarray}
\label{eq:cfir}
\mathcal{F}_{r}  & = & \displaystyle  
\int_0^{\infty}\frac{[\partial_z \int_0^{2\pi} 
p(r,\phi|z) \, d\phi ]^2}{\int_0^{2\pi} p(r,\phi|z)d\phi} \, rdr \, , \nonumber \\
& & \\
\mathcal{F}_{\phi} & = & \displaystyle 
\int_0^{2\pi}\frac{[\partial_z \int_0^{\infty}p(r,\phi|z) \, rdr ]^2}{\int_0^{\infty}p(r,\phi|z) \; rdr}d\phi \, , \nonumber 
\end{eqnarray}
respectively.  They quantify the information of  the radial and the azimuthal intensity components due to the $z$-dependence. For example, beam rotation only contributes to $\mathcal{F}_\phi $, whereas beam divergence contributes to $\mathcal{F}_r$.

For pure $|\mathrm{LG}_{pl} \rangle $ modes,  {$\Psi(r,\phi;z)=\mathrm{LG}_{pl}(r,\phi,z)$, the classical Fisher information can be obtained in a closed form. First notice the integrand of (\ref{eq:cfi}) is independent of $\phi$. Carrying out $z$-derivatives of Laguerre polynomials using the relation $\partial_z L_p^l(z)=-L_{p-1}^{l+1}(z)$, and changing the integration variable $r$ yields 
\begin{equation}
\mathcal{F}= \frac{4 p!}{(|l|+p)! }\left[\frac{\partial_z w(z)}{w(z)}\right]^2\int\limits_0^\infty e^{-t} t^{|l|}
\left[ 2t\, L_{p-1}^{|l|+1}(t)+(t-|l|-1)L_p^{|l|}(t) \right]^2 dt \, .
\end{equation}
We are thus left with evaluating six different integrals involving products of Laguerre polynomials. Compact expressions for each case can be obtained from the generalized orthogonality relation~\cite{Rassias:1992aa}
\begin{equation}
\fl
\int_0^\infty \rme^{-t} t^\mu L_p^l(t) L_{p'}^{l'}(t) \rmd t =
(-1)^{p+p'} \Gamma(\mu+1)\sum_{k=0}^{\min(p,p')}
{\mu-l \choose p-k}{\mu-l' \choose p'-k}{\mu+k \choose k} \, .
\end{equation}
Straightforward simplifications leave us with the final result 
}
\begin{equation}
\label{fish_intens}
\fl \mathcal{F}  =  4[ 2p ( p + |l|)+ 2p +|l| +1 ]
\left [ \frac{\partial_zw(z)}{w(z)} \right]^2 
 =  \frac{2p(p+|l|)+2p+|l|+1}{R(z)^2/4} \, .
\end{equation}
For the planes of maximal wavefront curvature $z_{\mathrm{opt}}= \pm z_{{\mathrm{R}}}$, where $R(z_{\mathrm{R}})^2= 4 z_{\mathrm{R}}^2$, the  quantum limit is saturated with intensity sensitive detection
\begin{equation}
\label{res4}
\mathcal{F}_{\mathrm{opt}} = \mathcal{Q} \, .
\end{equation}
Thus, for any pure LG beam and any beam waist location, two detection planes can be found, where complete information about the axial distance can be extracted with intensity-only detection. Hence, full potential of high-order vortex beams for axial metrology can be exploited with direct detection techniques. This generalizes the results  obtained for Gaussian beams\cite{Rehacek:2019aa}.

However, numerical analysis suggests that this result does not hold for superpositions of LG modes: a single transversal intensity scan is no longer optimal for any detector position $z$. In figure~\ref{fig:FIQl} we plot the classical Fisher information (normalized to the optimal QFI) for different vortex superpositions. First, notice that the optimal detector position is no longer at $z= \pm z_{\mathrm{R}}$,  as it was for pure LG beams. Second, the larger the angular momentum $l$ carried by the signal, the smaller portion of the total QFI can be extracted with intensity measurements.   

\begin{figure}[t]
\centering
\includegraphics[height=5.5cm]{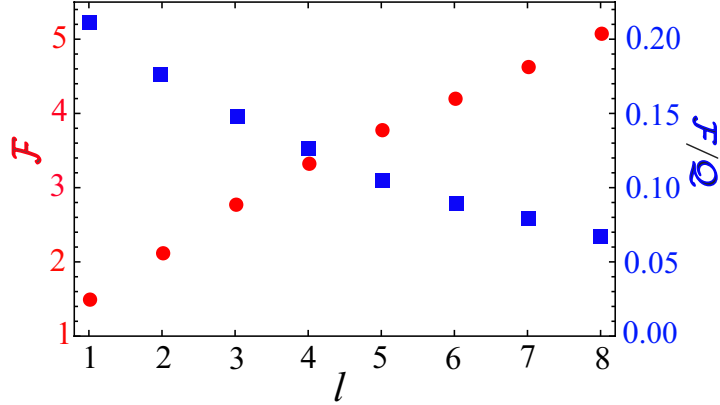}
\caption{Red dots represent Fisher information  about axial distance $z$ for optimally placed intensity detections for the superpositions $(|\mathrm{LG}_{0l} \rangle + | \mathrm{LG}_{00} \rangle )/\sqrt{2}$. Blue squares are the same Fisher information, but normalized to the optimal QFI.}  
\label{fig:maxFIl}
\end{figure}

\begin{figure}[b]
\centering
\includegraphics[height=5.5cm]{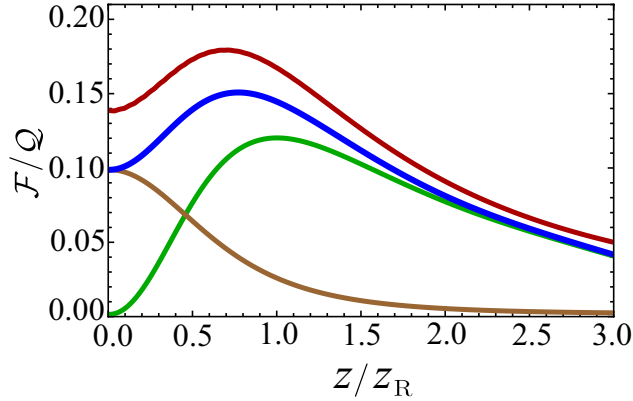}
\caption{Radial (green), azimuthal (brown) and total (red) Fisher information $\mathcal{F}$ about axial displacement as a function of  detector axial position $z$ (in units of $z_{\mathrm{R}}$) for the vortex superposition $(|\mathrm{LG}_{02} \rangle + \mathrm{LG}_{00})/\sqrt{2}$. The sum of radial and azimuthal components is also shown (blue). All quantities are normalized to the corresponding optimal QFI. Notice that for small (large) propagation distances, most FI comes from azimuthal (radial) intensity distribution. Some information is stored in correlations between those two marginal distributions, as can be seen from the gap between the blue and the red curves.}
\label{fig:FIrphi}
\end{figure}

This is further illustrated in figure~\ref{fig:maxFIl}. Locating intensity detectors at the optimal detection planes for every value of $l$, the classical Fisher information grows sublinearly with $l$, so that progressively smaller portion of quadratic QFI is available from intensity data and the gap between the quantum bound and the performance of the best estimator from intensity data worsens with $l$.

For example, for $l=2$,  only about $17\%$ of the  QFI is available from intensity scan at the optimal detector position, as we can see in figure~\ref{fig:FIrphi}. Of this, only a small fraction is due to changes in the azimuthal intensity profile, as arises, e.g., in rotations, except for very small propagation distances, where azimuthal profile accounts for up to $10\%$ of QFI. This ratio quickly approaches zero with increasing $l$. 

The information content of rotating beam intensity distributions is surprisingly low. As bad as it sounds, this is not a negative result, but encouraging news for quantum metrology. It highlights inadequacy of simple detection techniques in this particular metrological scenario, where intensity detection fails to reveal all potentially accessible information about the parameter of interest. The hidden Information has to be accessed with advanced detection techniques. Here a generic tool is  projecting/sorting the signal into optimal set of spatial modes derivable from the quantum detection theory. Hence, we uncover a huge potential of axial superlocalization based on spatial-mode projections applied to higher order vortex beam superpositions~\cite{Zhou:2019aa}.
 
\section{Concluding remarks}

We have established the ultimate quantum limits for axial localization using vortex beams. For pure LG beams, this limit is attained with an intensity scan with the detector located at one of two optimal planes. For superpositions of LG beams, in particular of those with intensity profiles rotating on defocusing, this is no longer true. This means that microscopy methods based on rotating vortex beams may benefit from replacing traditional intensity scans with advanced mode-sorting techniques.
 
We thank Robert W. Boyd and Aaron Z. Goldberg for helpful discussions. {We acknowledge financial support of the project ApresSF, under the QuantERA programme, which has received funding from the European Union's Horizon 2020 research and innovation programme, the program H2020 (project StormyTune), Grant Agency of the Czech Republic (Grant No.~18-04291S), the Palack\'y University (GrantNo. IGA\_PrF\_2020\_004), and the Spanish MINECO (Grant PGC2018-099183-B-I00).}

\newpage
	

\end{document}